\documentclass[%
superscriptaddress,
twocolumn,
 amsmath,amssymb,
 aps,
prx,
]{revtex4-2}

\usepackage{graphicx}
\usepackage{dcolumn}
\usepackage{bm}
\usepackage{braket}
\usepackage[dvipsnames]{xcolor}
\usepackage[colorlinks=true,linkcolor=blue,citecolor=blue, urlcolor=blue]{hyperref}
\usepackage{comment}

\newcommand{\sissa}{Scuola Internazionale Superiore di Studi Avanzati (SISSA), I-34136 Trieste, Italy} 
\newcommand{\epfl}{Theory and Simulation of Materials (THEOS), and National Centre for Computational Design and Discovery of Novel Materials (MARVEL), École polytechnique fédérale de Lausanne (EPFL), CH-1015 Lausanne, Switzerland}
\newcommand{\cam}{Theory of Condensed Matter, Cavendish Laboratory,University of Cambridge, Cambridge, CB3 0US, United Kingdom}
\newcommand{\paul}{Paul Scherrer Institut (PSI), CH-5232 Villigen, Switzerland}

\newcommand{\SMdiff}{S1}
\newcommand{\SMlw}{S2}

\begin{document}

\title{A Universal Diffusion Model for the Quantum Position Operator}

\author{Ransell D'Souza}
\email{rdsouza@sissa.it}
\affiliation{\sissa}
\author{Austin Zadoks}
\affiliation{\paul}
\author{Riccardo Dal Molin}
\affiliation{\sissa}

\author{Nataliia Manko}
\affiliation{\sissa}

\author{Junfeng Qiao}
\affiliation{\epfl}
\affiliation{\paul}

\author{Nicola Marzari}
\affiliation{\epfl}
\affiliation{\paul}
\affiliation{\cam}

\author{Antimo Marrazzo}
\email{amarrazzo@sissa.it}
\affiliation{\sissa}

\date{\today}

\begin{abstract}
The quantum position operator and its response to perturbations underpin the functional properties of insulating materials. In the modern theory of polarization, Wannier charge centres encode Berry-phase polarization, while their structural derivatives determine key response functions. Here we introduce a universal diffusion model based on E(3)-equivariant graph neural networks for moments of the position operator, with the first moment defining Wannier centres and the second central moment their real-space localization. We demonstrate the framework for liquid water, amorphous silicon and crystalline BaTiO$_3$, and train a universal model that generalizes to 100 unseen crystals spanning a 67-species chemical space. The predicted centres reproduce infrared spectra in large liquid and amorphous cells beyond routine first-principles reach and enable finite-difference calculations of Born effective charges and piezoelectric tensors. This framework provides a scalable route to electronic polarization and localization in complex materials, supporting high-throughput screening and finite-temperature simulations.
\end{abstract}

\maketitle

The quantum position operator and the corresponding expectation values and derivatives are central to insulating matter and to technologies that exploit it. Electronic polarization describes how an insulator responds to an electric field, how charge redistributes during atomic motion, and how mechanical deformation couples to electrical signals. These effects are particularly strong in polar materials, where long-range electrostatic interactions shape lattice dynamics, phase stability and functional behaviour~\cite{grasselli_arxiv_2026}. They underpin capacitors, sensors, actuators, transducers, energy harvesters, infrared optics and microelectronics. Polarization also relates to microscopic spectroscopic observables: in liquids and disordered solids, the time correlation function of the total polarization can be used to determine the infrared absorption spectrum, making polarization fluctuations a direct probe of local structure, bonding and dynamics.

\begin{figure*}[!]
    \centering
    \includegraphics[width=1\linewidth]{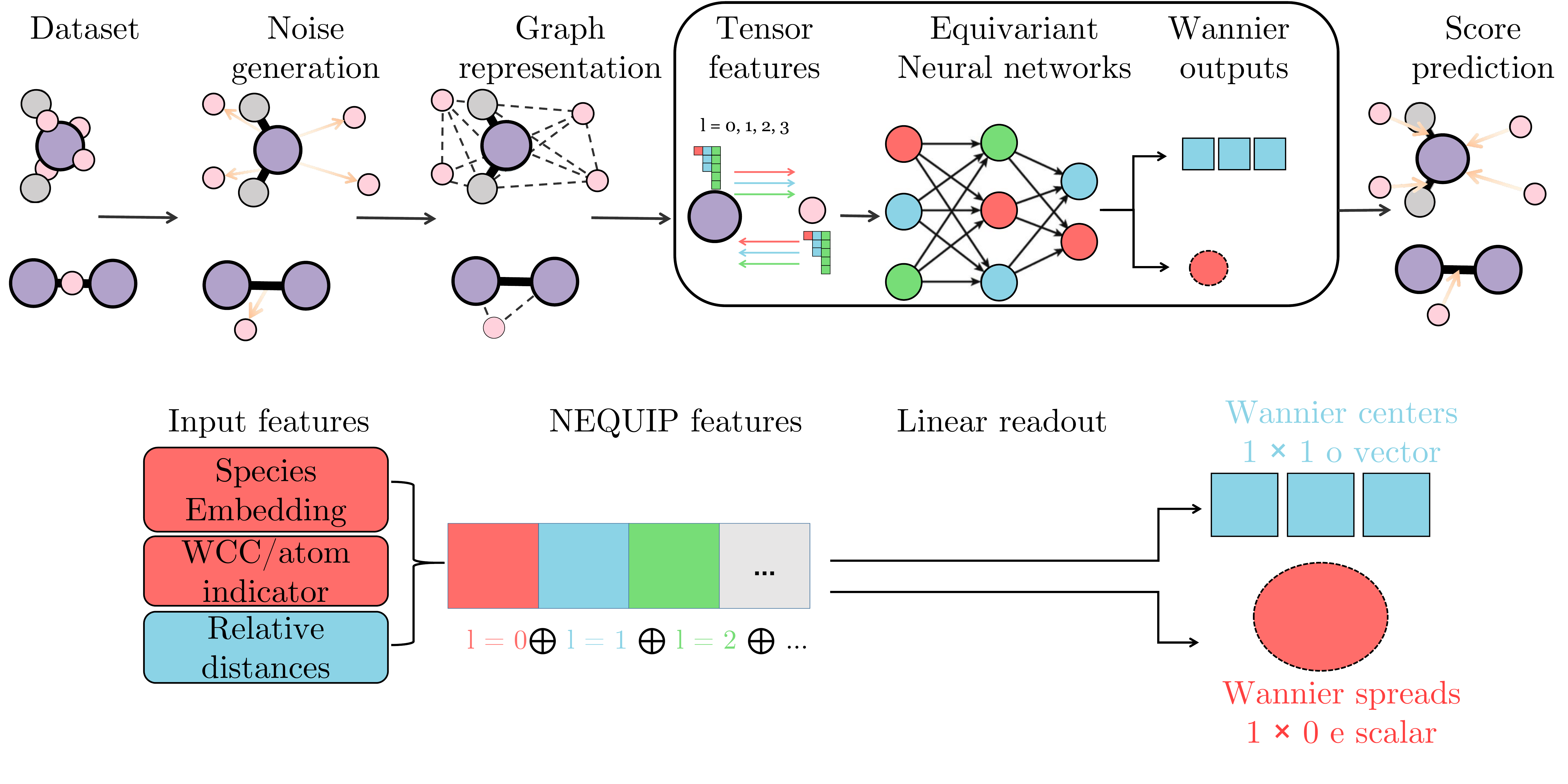}
    \caption{\textbf{\textsc{WannDiffuse} workflow:} (Upper  panel) Schematic of the data generation and diffusion process. Structures obtained from first-principles calculations are represented as atomic graphs containing atoms and Wannier charge centres (WCCs) as graph nodes. Noise is added to the WCC positions to generate perturbed configurations, from which the model learns the score function that guides randomly initialized WCCs towards their target positions.
    (Lower panel) Expanded view of the equivariant neural network. Node inputs comprise species embeddings, a WCC/atom indicator and relative distances, which are transformed into equivariant NequIP features with angular orders ($l = 0, 1, 2, \cdots $). Linear readouts produce a $1 \times 1o$ vector for each WCC and a $1 \times 0e$ scalar for its Wannier spread. The predicted scores are subsequently used to denoise the WCC positions.
    }
    \label{fig:workflow}
\end{figure*}

The derivatives of polarization encode equally important properties. The Born effective charge tensor measures the change in macroscopic polarization induced by an atomic displacement, and therefore represents the coupling between lattice vibrations and macroscopic electric fields. The piezoelectric tensor measures the change in polarization induced by strain, or equivalently the electrical response to mechanical deformation. Together, polarization, Born effective charges and piezoelectric tensors provide a compact description of dielectric and electromechanical behaviour. Accurately predicting these quantities at low computational cost and complexity, across finite temperatures and the length and time scales relevant to complex materials, remains a central challenge.

In periodic boundary conditions, electric polarization is not a conventional dipole moment per unit volume. Instead, the modern theory of polarization expresses it as a Berry phase of the occupied electronic manifold~\cite{Resta1992,Resta1994,KSV1993,KSV1993bis,Vanderbilt2018}. For a crystalline insulator~\footnote{A factor of 2 must be accounted for two-fold degenerate state in spin-unpolarized collinear calculations.}
\begin{equation}
\mathbf{P}
= \frac{e}{V_c}\sum_I Z_I \mathbf{R}_I
- \frac{f e}{(2\pi)^3}\sum_{n}^{\mathrm{occ}} \int_{\mathrm{BZ}} d\mathbf{k}\,
\langle u_{n\mathbf{k}} | i\nabla_{\mathbf{k}} | u_{n\mathbf{k}} \rangle ,
\label{eq:berry_polarization}
\end{equation}
where $V_c$ is the cell volume, $f=2$ in spin-degenerate bands, \(Z_I\) and \(\mathbf{R}_I\) are the ionic charges and positions, and \(u_{n\mathbf{k}}\) are the cell-periodic occupied Bloch states. Equation~\eqref{eq:berry_polarization} also exposes a fundamental subtlety, by now well understood: polarization is defined only modulo a polarization quantum, \(e\mathbf{R}/V_c\), with \(\mathbf{R}\) a lattice vector, or, equivalently polarization is a lattice. Observable quantities are therefore polarization differences or derivatives.

The same Berry-phase polarization has an intuitive real-space representation in terms of Wannier functions. For an isolated occupied manifold, Wannier functions ~\cite{wannier37} are localized orbitals obtained from the Fourier transform of unitary rotations of Bloch eigenstates,
\begin{equation}
\ket{\mathbf{R}j} = \frac{V_c}{(2\pi)^3}\int_{BZ} d\mathbf{k}e^{-i\mathbf{k}\cdot\mathbf{R}}\ket{\psi_{j\mathbf{k}}^W}
\label{eq:wannier_definition}
\end{equation}
where  $ \ket{\psi_{j\mathbf{k}}^W}= \sum_{n=1}^{J}U_{\mathbf{k},nj}\ket{\psi_{n\mathbf{k}}} $
is chosen to enforce a smooth and periodic gauge within the occupied subspace over the entire Brillouin zone. The centre of each Wannier function is 
\begin{equation}
\bar{\mathbf{r}}_j = \langle \mathbf{0}j | \mathbf{r} | \mathbf{0}j \rangle 
\label{eq:wannier_center}
\end{equation}
and so the polarization can be written as
\begin{equation}
\mathbf{P}
= \frac{e}{V_c}\left(\sum_I Z_I\mathbf{R}_I - f\sum_{j}^{\mathrm{occ}} \bar{\mathbf{r}}_j\right)
\quad \mathrm{mod}\; \frac{e\mathbf{R}}{V_c}.
\label{eq:wannier_polarization}
\end{equation}
Thus the sum of these centres, known as Wannier charge centres (WCCs), is gauge invariant modulo the polarization quantum, even though individual Wannier functions are gauge dependent.

A particularly useful localization criterion is provided by maximally localized Wannier functions (MLWFs)~\cite{marzari_prb_1997,marzari_rmp_2012,marrazzo_rmp_2024}. They are defined as the gauge that minimizes the total quadratic spread of the position operator
\begin{equation}
\Omega = \sum_{j=1}^{J}\left[\braket{\mathbf{0}j|r^2|\mathbf{0}j}-|\braket{\mathbf{0}j|\mathbf{r}|\mathbf{0}j}|^2\right]
\label{eq:mlwf_spread}
\end{equation}
with respect to the unitary matrices $U_{\mathbf{k},nj}$. This localization criterion yields compact orbitals and charge centres that provide chemically interpretable descriptions of covalent bonds, lone pairs and charge transfer~\cite{marzari_rmp_2012,marrazzo_rmp_2024}.

Despite their success, first-principles calculations, including the corresponding post-processing MLWF analysis, remain too demanding for many phenomena of substantial interest. Infrared spectra of liquids require large simulation cells and long trajectories, while amorphous materials demand extensive configurational sampling. Ferroelectric and piezoelectric materials exhibit anharmonic dynamics, finite-temperature domain fluctuations and field-induced responses over scales far beyond routine density functional theory (DFT). Machine learning interatomic potentials (MLIPs) have greatly extended first-principles-quality simulations of atomic structure and dynamics~\cite{BP2007,Deringer2019}, but they generally do not explicitly describe the electronic degrees of freedom needed to compute polarization and its derivatives.

Electronic-structure methods based on DFT, including density-functional perturbation theory (DFPT)~\cite{baroni_rmp_2001}, Berry-phase theory~\cite{Vanderbilt2018} and approaches based on maximally localized Wannier functions (MLWFs), provide a comprehensive framework for computing polarization and dielectric response in insulating materials~\cite{Gonze1997,Nunes2001,Souza2002}. Derivatives of the polarization can be evaluated either through DFPT or by finite differences (FD). DFPT can treat monochromatic perturbations directly within the primitive cell, whereas FD approaches often require larger supercells but offer the advantage of a simpler and more general numerical implementation that is readily applicable across different exchange-correlation functionals. FD schemes may be formulated starting from the electric-enthalpy functional~\footnote{We note that the electric-enthalpy formalism is required only when finite macroscopic electric fields are applied under periodic boundary conditions, and that it entails several technical complications. In particular, the corresponding functional is not globally bounded from below, and stable local minima exist only below a critical electric field that decreases as the reciprocal-space $k$-point sampling is refined.}, with polarization obtained as the derivative of the energy with respect to the electric field, or directly from the polarization itself, evaluated either through the Berry phase or from the sum of WCCs. Despite these complementary formulations, all approaches remain computationally demanding for complex systems. DFPT requires the self-consistent solution of the Sternheimer equation, while Wannierization entails an iterative gauge optimization that can become increasingly challenging for large occupied manifolds; in practice, however, the overall cost is typically dominated by the underlying self-consistent first-principles calculation. These computational requirements severely limit the evaluation of polarization and its derivatives in large or structurally complex systems, hindering high-throughput materials screening and design as well as large-scale molecular-dynamics (MD) simulations. Similar limitations affect the characterization of electron localization through the quadratic spread of the position operator. Extending calculations of electron localization to complex, dynamical environments would provide valuable support for studies ranging from solvated electrons~\cite{gao2024,gao_arxiv_2025} to polaron formation and dynamics~\cite{franc_nrm_2021,franc_prl_2025}.

Similar to the field of MLIPs, ML seems a promising route to accelerate calculations of electronic-structure properties related to the position operator. Recent developments focused on learning response operators and tensorial observables, 
including a unified differentiable framework that jointly predicts electric enthalpy, polarization, Born charges, and polarizability within a single equivariant model \cite{Falletta2025, Martin2026},
operator quantum machine learning (OQML) \cite{Christensen2019}, symmetry-adapted Gaussian process regression (SA-GPR) and SOAP-based kernel methods for tensorial response \cite{Grisafi18, Wilkins2019, Grisafi2019}, neural networks with explicit charge and dipole degrees of freedom such as PhysNet \cite{Unke2019} and MuML \cite{Veit2020}, and field-aware equivariant architectures such as FieldSchNet \cite{Gastegger2021}. 
Analogous approaches have been extended to crystalline materials to predict dielectric constants and DFPT-derived tensors directly from structure~ \cite{Morita2020, Lou2025}, while electrically aware MLIPs have incorporated latent charges, dipoles, and Wannier-center proxies to recover polarization and long-range electrostatics in atomistic simulations~\cite{Zhang2020, Zhang2022}. 
In particular, DeepWannier and the Deep Potential Long-Range (DPLR) framework demonstrated that ML models can predict Wannier \textit{centroids} with sufficient fidelity to reproduce polarization and electrostatic interactions in MD~\cite{Zhang2020, Zhang2022, gao_arxiv_2025}. The Wannier-function approach to polarization and electrostatics offers the advantage of a strictly local real-space framework, in which polarization can be expressed as a sum of local contributions while simultaneously providing access to other relevant quantities, such as degree of electron localization. 

The only existing approaches that explicitly predict WCCs are atom-centred models that either predict Wannier centroids~\cite{Zhang2020} or rely on predefined assignments of WCCs to specific atoms~\cite{gao2024}. These approaches therefore impose an atom-centred representation of the electronic degrees of freedom, limiting their ability to treat WCCs as an intrinsically unordered set and, in turn, hindering general applicability and the development of universal or foundation models. This makes WCC prediction particularly well suited to a generative diffusion formulation: direct regression requires an explicit assignment or ordering of the target centres, which can become ambiguous as bonding environments change or when WCCs occupy interstitial positions. By generating WCCs iteratively as an unordered set, diffusion avoids this assignment problem and naturally accommodates non-atom-centred electronic localization.

Here we introduce \textsc{WannDiffuse}, a universal diffusion model for the quantum position operator based on predicting the centres and real-space localization of MLWFs. Our framework combines \(E(3)\)-equivariant graph neural networks based on NequIP~\cite{Satorras2021,Batzner2022} with a generative diffusion process~\cite{song2019generative, Ho2020,song2021scorebased,Hoogeboom2022,Zeni2025} to predict WCCs and spreads directly from atomic configurations. In contrast to previously-reported atom-centred approaches, our diffusion framework represents WCCs as an additional atomic species and learns to generate them as an unordered, structure-conditioned set, enabling transferable predictions across diverse chemistries, bonding environments and phases.
Rather than learning a material-specific charge model, the model learns a transferable distribution of localized electronic charge centres conditioned on structure. This enables the reconstruction of Berry-phase polarization and its fluctuations, as well as Born effective charges and piezoelectric tensors obtained by FD. We demonstrate wide applicability across liquid water, amorphous silicon, crystalline BaTiO$_3$ and 100 unseen crystalline materials spanning a 67-species chemical space. 

While in this work we focus on the first- (polarization) and second-order (spread) moments of the position operator, the framework can accommodate arbitrary $n$-th-order moment tensors of the position operator if reference values are available.

\section*{Results}
\begin{figure}[!]
\includegraphics[width=1\linewidth]{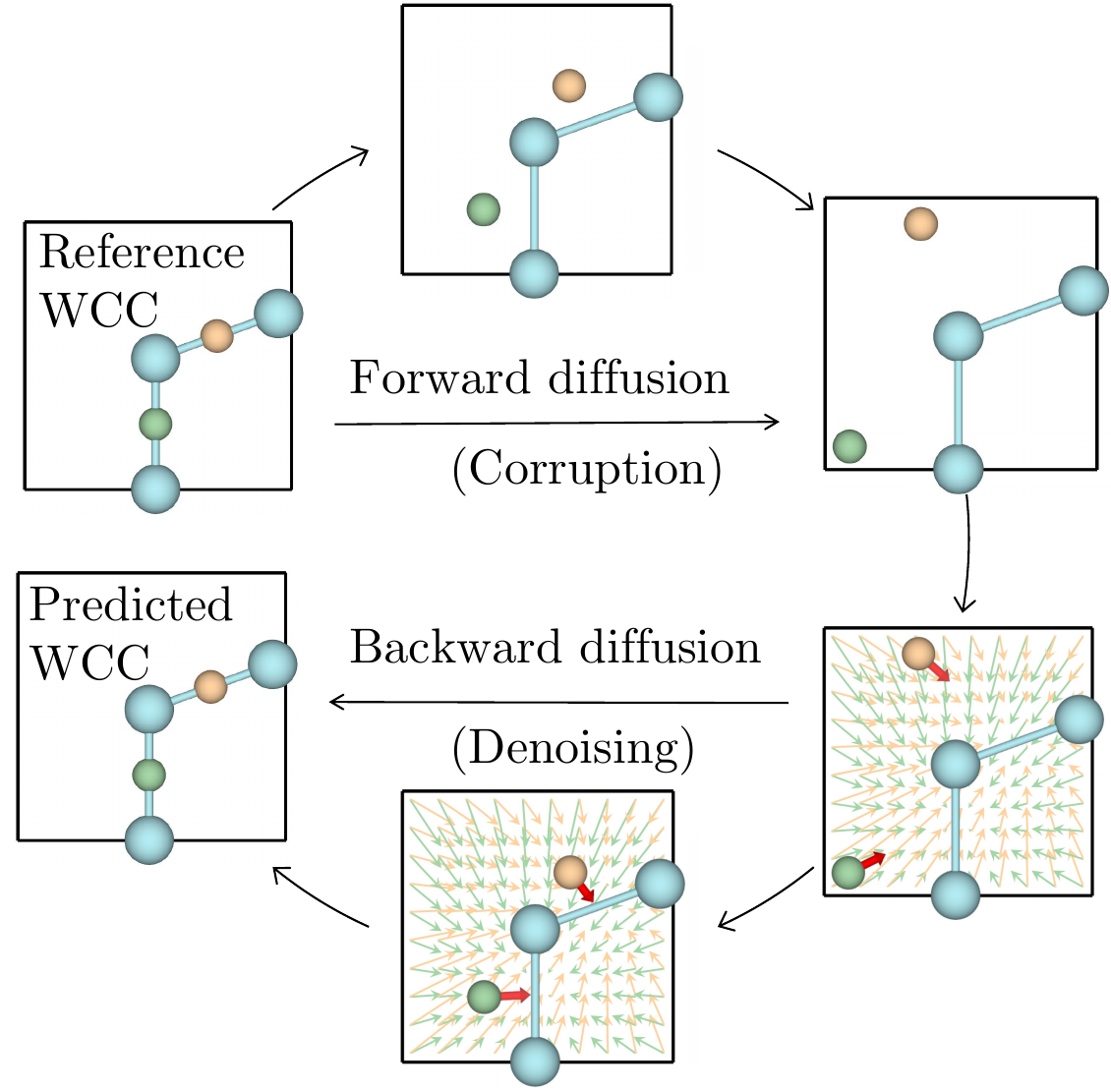}%
\caption{\label{fig:diff_toy} \textbf{Visualizing the diffusion framework for Wannier-centre prediction:} Schematic representation of the forward and reverse diffusion processes underlying \textsc{WannDiffuse}. The initial configuration (top left) represents the reference Wannier-centre distribution sampled from the dataset. During the forward diffusion process (top row), progressively increasing Gaussian noise is added to the WCC, transforming the physical electronic structure into a perturbed distribution. The reverse diffusion process (bottom row) reconstructs the reference configuration by iteratively following the learned score field (arrows), which indicates the direction of the denoising trajectory for each WCC. The coloured arrows correspond to the predicted displacement directions of WCCs towards their reference positions. After successive reverse diffusion steps, the generated WCC configuration should converge to the true distribution (bottom left).}
\end{figure}
\subsection{\textsc{WannDiffuse} framework}
Figure~\ref{fig:workflow} summarizes the \textsc{WannDiffuse} pipeline. First-principles reference data are converted into a graph in which both atoms and WCC are nodes, distinguished by a binary species-type feature.

\paragraph{Equivariant graph neural networks}
Graph connectivity is constructed under periodic boundary conditions, and node and edge features are passed through an $E(3)$-equivariant message-passing network, which outputs, for every node tagged as a WCC, a displacement vector and an associated scalar spread (details in Methods Sec.~\ref{sec:gnn}).

\paragraph{Diffusion model}
The forward process generates training pairs by adding Gaussian noise to the true WCC positions (see Supplementary Material Sec. \SMdiff .1 for a derivation). The network is trained to map each noised graph back to the displacement that recovers the clean configuration. During inference, WCCs are initialized at arbitrary positions and iteratively refined by repeated application of the trained denoiser. After a user-specified number of iterations, the predicted centres converge to the true Wannier-centre positions.

Figure~\ref{fig:diff_toy} demonstrates this reverse process on a minimal toy system, a triatomic arrangement with bond-centred WCCs, chosen because the ground truth is trivial to verify by inspection. 
WCCs initialised at random positions converge to the true bond-midpoint locations within the full set of denoising steps, with the predicted score field pointing toward the true centres at every point in the surrounding space, not only along the trajectory actually taken. 
Convergence is monotonic in the displacement from the true position: early, high-noise steps produce large corrective moves, and later, low-noise steps produce small refinements.

Because WCC generation is formulated in terms of local atomic environments, the computational cost of training depends primarily on the number of configurations and diffusion iterations rather than on the total size of the simulated system. 
Generating WCCs for a trajectory containing (N) configurations therefore requires (N $\times$ T) forward passes of the neural network, where T is the number of diffusion iterations (e.g., T=15). 
For a fixed number of training configurations, increasing the number of diffusion timesteps can further improve the accuracy of the model.
For a fixed denoising schedule, the number of network evaluations is independent of system size. At fixed density and cutoff, the local message-passing architecture is expected to give an approximately linear increase in computational cost with the number of atoms and Wannier centres.
This enables the 512-molecule water trajectory (Sec.~\ref{sec:liq_water}) and the 5,000-frame amorphous-silicon trajectory (Sec.~\ref{sec:aSi}) to be processed using models trained on substantially smaller reference calculations. The universal model further demonstrates transferability across chemically and structurally diverse insulating materials.

\begin{figure*}[t!]
\includegraphics[width=1\linewidth]{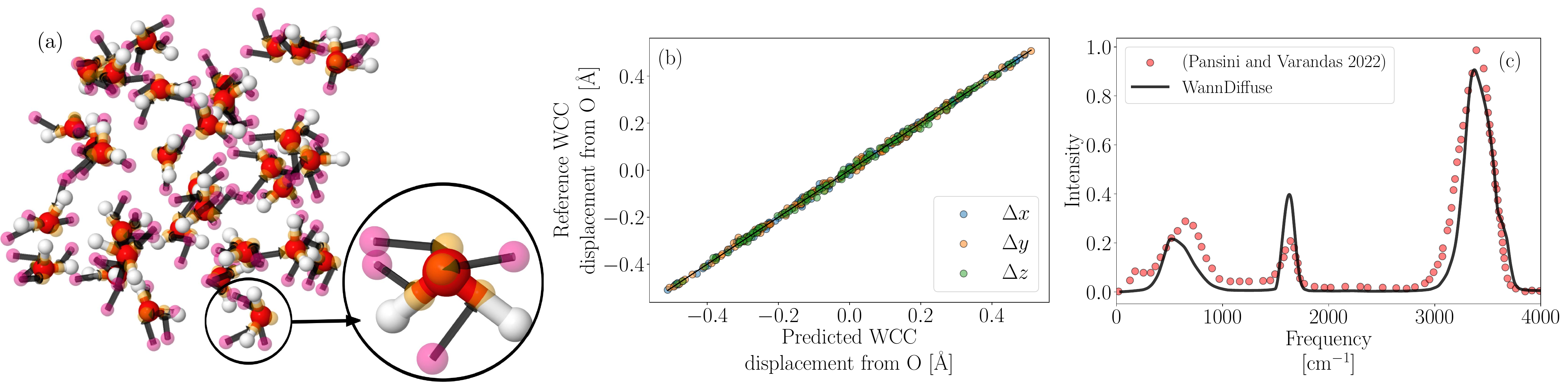}
\caption{\label{fig:liq_water_res} \textbf{Liquid water:}
(a) Illustration of the diffusion process in liquid water. Pink spheres denote randomly initialized WCCs, while black lines indicate the score predicted by the model. Orange spheres represent the updated positions at the next diffusion step, corresponding to the predicted WCCs. The inset shows a magnified view of a single water molecule.
(b) Scatter plot comparing true and predicted WCC positions for a validation frame.
(c) Infrared (IR) spectrum of liquid water computed for a larger system of 512 water molecules (H$_2$O)$_{512}$, trained on 800 samples of uncorrelated (H$_2$O)$_{32}$ configurations. The experimental IR spectrum of liquid water from Ref. \cite{PANSINI2022} is shown for comparison.}
\end{figure*}

\subsection{Liquid water}\label{sec:liq_water}
For the liquid-water model, we train \textsc{WannDiffuse} on 800 structures from the 32-molecule path-integral MD (PIMD) dataset of Grisafi et al. \cite{Grisafi18}. Further details on the training procedure and dataset are provided in the Methods section and Supplementary Material (SM). Scaling the training set from 1 to 800 configurations reveals rapid initial improvement followed by saturation, with the learning-curve exponent decreasing from -0.451 to -0.101 beyond approximately 250 configurations \cite{supp}. This behaviour indicates that \textsc{WannDiffuse} efficiently captures the dominant structural–electronic correlations of liquid water with relatively small training sets.
The observed scaling behaviour is consistent with the regime reported for NequIP \cite{Batzner2022}, indicating that the model follows the standard learning-curve trends expected for equivariant NNIPs, while operating within the same data-efficiency range as established architectures for atomistic machine learning.

We further evaluate the ability of \textsc{WannDiffuse} to reproduce dynamical electronic properties by predicting Wannier-centre trajectories for a 512-molecule liquid-water supercell generated from {\it ab initio} MD~\cite{Pabst2025}. The resulting dipole autocorrelation function yields an infrared spectrum that accurately reproduces the characteristic vibrational signatures of liquid water, including the hydrogen-bond-broadened O–H stretching region, the H–O–H bending mode, and the low-frequency librational band. The predicted vibrational frequencies agree with experiment to within a few $cm^{-1}$ for the stretching and bending modes~\cite{PANSINI2022}, with the corresponding frequencies reported in Table S1 of the SM. The good agreement demonstrates that the learned Wannier-centre dynamics accurately capture the fluctuating electronic polarization required for the prediction of spectroscopic quantities.
A full comparison with DeepWannier can be found in Supplementary Section \SMlw .1

\subsection{Amorphous silicon}\label{sec:aSi}
\begin{figure*}[!]
\includegraphics[width=1\linewidth]{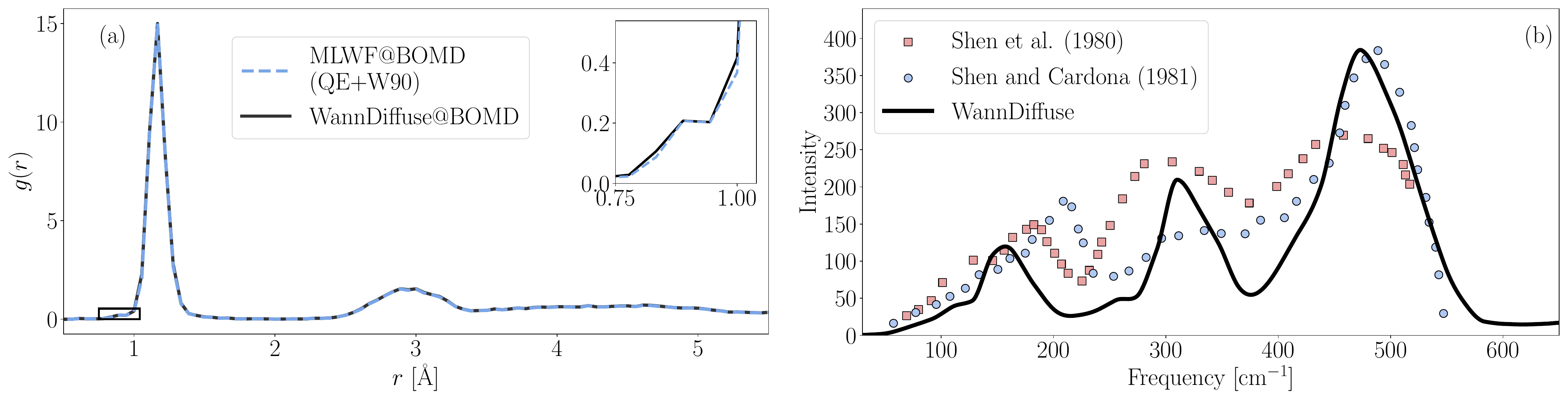}
\caption{\label{fig:aSi_res} \textbf{Amorphous Silicon:}
(a) Radial distribution functions (RDFs) for Si–WCC pairs. The Si–WCC RDF is computed from 64-atom uncorrelated Born–Oppenheimer molecular dynamics (BOMD) snapshots, with the WCCs obtained using \textsc{WannDiffuse} (solid black) and \textsc{Quantum Espresso} + \textsc{Wannier90} (dashed blue) for each configuration.
(b) Infrared (IR) spectrum obtained from the dipole polarization computed using the predicted WCCs along a 512-atom MACE MD trajectory. The experimental IR spectra from references \cite{Shen1980, Shen1981} are shown for comparison.}
\end{figure*}

For amorphous silicon (a-Si), we train a separate \textsc{WannDiffuse} model on 120 structures generated using AiiDA-TrainsPot~\cite{Bidoggia2026}, an automated AiiDA~\cite{Pizzi16,Huber20} workflow for dataset augmentation and NNIP training, sampled over 15 diffusion timesteps. Further details on the dataset-augmentation and labeling protocol are given in Methods.
The Si-Si radial distribution function, computed from the Born–Oppenheimer molecular dynamics (BOMD) trajectory, reproduces the characteristic short-range structural features of amorphous silicon, including the first and second coordination shells near 2.4 \AA~and 3.0 \AA, in agreement with the 64-atom reference data of Silvestrelli et al.~\cite{SILVESTRELLI19987, supp}.  The broader distribution at longer distances reflects the absence of crystalline order.
In Fig. \ref{fig:aSi_res} (a) the Si–WCC RDF reproduces the expected coordination geometry, with some structure for distances smaller than 1 \AA, hence very close to the Si ions. Regarding the first small peak, the \textsc{WannDiffuse} results closely match the \textsc{Wannier90} calculations performed on the same Born--Oppenheimer MD (BOMD) trajectory, while both deviate slightly from Ref.~\cite{SILVESTRELLI19987}, where this feature was interpreted as a clear signature of defects in the system. We attribute this small discrepancy to differences in the underlying trajectories: the trajectory used in Ref.~\cite{SILVESTRELLI19987} was generated using Car--Parrinello MD and likely employed different DFT settings and simulation parameters.

To evaluate scalability beyond the training regime, \textsc{WannDiffuse} was applied to a 512-atom a-Si supercell sampled over 5,000 MD frames generated using a MACE interatomic potential \cite{Batatia2022mace, Batatia2022Design} trained by us (see Methods). The model easily predicts the thousands of WCCs that correspond to each configuration, enabling converged structural statistics for system sizes inaccessible to direct first-principles calculations.

The predicted WCC dynamics further enable accurate calculation of the infrared response of amorphous silicon.  Fig. \ref{fig:aSi_res} (b) shows the IR spectrum of a-Si. The resulting spectrum reproduces the three dominant experimental vibrational features reported by Shen et al. \cite{Shen1980, Shen1981}, demonstrating that \textsc{WannDiffuse} captures the coupling between atomic motion and electronic polarisation in a disordered solid. A small systematic redshift relative to experiment is observed; however, the same deviation is present in the vibrational density of states obtained directly from the underlying atomic trajectory, indicating that the discrepancy originates from the atomic dynamics rather than from the Wannier-centre prediction. The spectrum obtained from the 512-atom cell also agrees with previous 64-atom calculations by Debernardi et al. \cite{Debernardi1997}, while benefiting from improved configurational sampling that reduces finite-size artefacts and reproduces the broad spectral features characteristic of amorphous silicon.

\subsection{Barium titanate}

To evaluate whether \textsc{WannDiffuse} can capture the response of a polar functional material, we train a separate model for barium titanate (BaTiO$_3$) using 50 initial structures sampled over 50 diffusion timesteps (see Methods).
We predict the Born effective charges and piezoelectric tensors of cubic and tetragonal BaTiO$_3$.
The Born effective charges obtained from the predicted WCCs closely reproduce established first-principles benchmarks, with deviations of approximately $0.1e$ for all atomic species relative to previous calculations~\cite{Marzari1998,Ghosez1995,Zhong1994} (Table~\ref{tab:bec}). In particular, the anomalously large effective charges of Ti and O, arising from strong Ti--O hybridisation and playing a central role in the ferroelectric response of BaTiO$_3$, are accurately recovered. The predicted spontaneous polarisation of tetragonal BaTiO$_3$, $\Delta P = 0.288\ \mathrm{C~m^{-2}}$, is in close agreement with experimental measurements ($0.26\ \mathrm{C~m^{-2}}$) \cite{Megaw1962,Wieder1955} and previous first-principles calculations \cite{Marzari1998,Ghosez1995,Zhong1994}.

\begin{table}[h]\label{tab:bec}
\centering
\caption{Born effective charges (in units of $e$) for cubic BaTiO$_3$.
$\mathrm{O}_\parallel$ and $\mathrm{O}_\perp$ denote oxygen atoms displaced
parallel and perpendicular to the Ti--O bond, respectively.}
\label{tab:bec}
\begin{ruledtabular}
\begin{tabular}{lcccc}
 & Ba & Ti & O$_\parallel$ & O$_\perp$ \\
\hline
\textsc{WannDiffuse} (FD)                                  & 2.758 & 7.136 & $-5.706$ & $-2.094$ \\
DFPT & 2.759  & 7.362 & $-5.823$ & $-2.149$ \\
Marzari \textit{et al.}~\cite{Marzari1998} & 2.65  & 7.33  & $-5.87$  & $-2.06$  \\
Ghosez \textit{et al.}~\cite{Ghosez1995}   & 2.77  & 7.25  & $-5.71$  & $-2.15$  \\
Zhong \textit{et al.}~\cite{Zhong1994}     & 2.75  & 7.16  & $-5.69$  & $-2.11$  \\
\end{tabular}
\end{ruledtabular}
\end{table}

Beyond static polarization, \textsc{WannDiffuse} also reproduces the clamped-ion piezoelectric response of tetragonal BaTiO$_3$. The predicted coefficients, $e_{31}=0.2361$, $e_{33}=-0.400$, and $e_{15}=0.1534\ \mathrm{Cm^{-2}}$, closely match our own Berry-phase FD calculations as well as first-principles results from the literature that were obtained using a range of exchange-correlation functionals~\cite{Rusevich2017}; see Table~\ref{tab:piezo} for the full comparison.

\begin{table}[h]
\centering
\caption{Clamped-ion piezoelectric tensor coefficients (in $\mathrm{C\,m^{-2}}$) of tetragonal BaTiO$_3$.}
\label{tab:piezo}
\begin{ruledtabular}
\begin{tabular}{lccc}
 & $e_{31}$ & $e_{33}$ & $e_{15}$ \\
\hline
\textsc{WannDiffuse} (FD)        & 0.2361  & $-$0.400  & 0.15341  \\
Berry-phase (FD)         & 0.2364  & $-$0.404  & 0.15263  \\
WCGGA \cite{Rusevich2017}  & 0.23  & $-0.399$ & 0.148 \\
B1WC   \cite{Rusevich2017}  & 0.233 & $-0.403$ & 0.152 \\
PBE0   \cite{Rusevich2017}  & 0.274 & $-0.471$ & 0.199 \\
\end{tabular}
\end{ruledtabular}
\end{table}

\subsection{Universal model}

\begin{figure*}[!]
    \centering
    \includegraphics[width=0.8\linewidth]{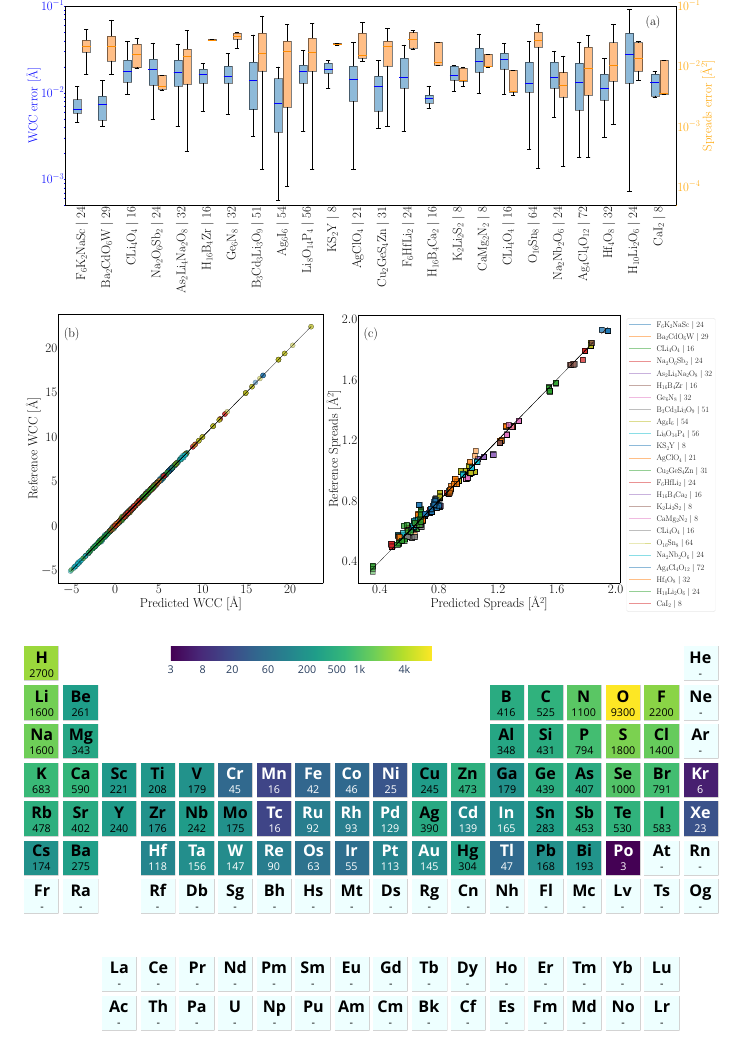}
    \caption{\textbf{Validation error of the universal diffusion
    model across unseen materials:} Performance is evaluated on 100 insulating materials excluded from training, with all 67 constituent chemical species represented in the training set. (a) Distribution of RMSE errors for the predicted WCCs and spreads for each composition. Each box plot shows the deviation between predicted and reference WCCs (blue) and spreads (orange); the box denotes the interquartile range (25th-75th percentiles), the horizontal line the median, and the whiskers indicate the most extreme values. The number following the vertical separator in each x-axis label denotes the number of Wannier functions. (b,c) Parity plots comparing predicted and reference WCCs (b) and spreads (c). (Lower panel) Heat map of the periodic table showing the total number of elemental occurrences in the dataset used to train the universal model.}
    \label{fig:error_metrics_universal}
\end{figure*}

A universal model is trained on a dataset of 2,807 crystalline insulators spanning 67 chemical species. Further details on the training procedure and dataset are provided in the Methods section.

To assess generalization beyond the structural environments encountered during training, we validate the model on 100 previously unseen insulating materials. The test set is constructed such that all constituent chemical species are represented in the training data, while no crystal structure overlaps with the training set.

As shown in Fig.~\ref{fig:error_metrics_universal}, the model exhibits consistently low errors across the hold-out set for both WCC and spreads. The parity plots further show close correspondence between predicted and reference values over the range encountered in the test set.
Quantitatively, the RMSE over the complete validation set is 0.09 \AA~for WCCs and 0.02 \AA$^2$ for spreads, demonstrating the overall accuracy of the model across both target properties.

\subsubsection{A focus on monolayer MoS$_2$}

In honeycomb lattices, such as those of 1H transition-metal dichalcogenides (TMDs), the bulk polarization is quantized into discrete classes indexed by $m\in{0,1,2}$ and protected by $C_3$ rotational symmetry~\cite{fang_prb_2012,jadaun_prb_2013,Gibertini_natcomm_2014,gibertini_nanolett_2015}. This non-trivial topology is reflected in the arrangement of the WCCs, which cannot all be continuously mapped onto atomic sites without either closing the energy gap or breaking $C_3$ symmetry. In monolayer MoS$_2$, eight WCCs associated with Mo--S bonding and sulfur lone-pair states localize around the sulfur atoms and along the Mo--S bonds, whereas an additional centre, predominantly derived from the Mo $d_{z^2}$ orbital, resides at the centre of the hexagonal ring. This interstitial WCC provides a real-space manifestation of the non-trivial topological polarization class and cannot be naturally assigned to any individual atom.
The universal \textsc{WannDiffuse} model accurately predicts all nine WCCs, including the interstitial centre, in excellent agreement with reference \textsc{Quantum Espresso}+\textsc{Wannier90} calculations (Fig.~\ref{fig:mos2}).

To determine whether this prediction arises from genuine transferability or chemical memorisation, we perform a controlled training-set ablation study and report the results in Fig.~\ref{fig:mos2}. A model trained without TMDs predicts WCC incorrectly, collapsing them towards atomic positions with a mean error of 0.501 Å, indicating that the characteristic non-atomic electronic localization is absent from the learned representation. Including other TMDs while excluding compounds containing Mo and S reduces the error to 0.325 Å; still, the interstitial and bonding WCC remain incorrectly positioned, demonstrating that structural similarity alone is insufficient if the considered chemical species are completely absent from the training dataset. 
In contrast, a model trained on Mo- and S-containing materials while excluding monolayer MoS$_2$ 
itself achieves an RMSE error of only 
$1.63\times10^{-2}$ \AA~and reproduces the WCC arrangement with high accuracy. These results demonstrate that \textsc{WannDiffuse} learns transferable chemical and structural representations of electronic localization, enabling it to exploit both chemical identity and local atomic environments to accurately predict WCC configurations in unseen materials, including topologically non-trivial cases, provided related chemical species and structural motifs are represented in the training data.

\begin{figure*}[!]\label{fig:mos2}
    \centering
    \includegraphics[width=\linewidth]{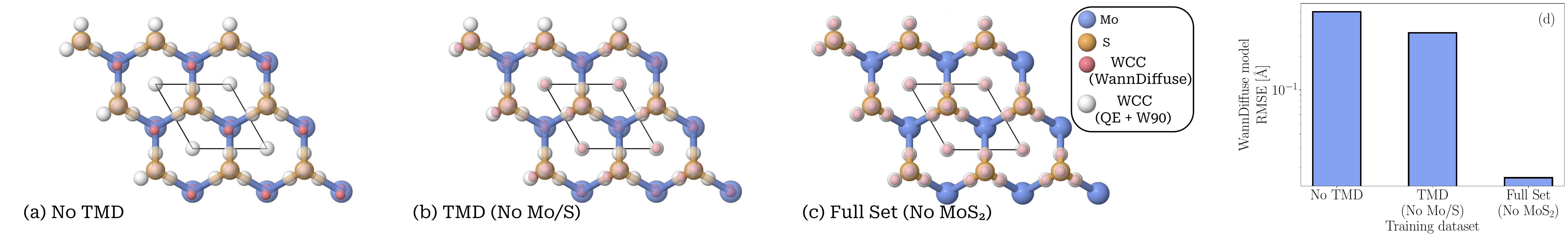}
    \caption{\textbf{Chemical transferability study for monolayer MoS$_2$}:
    Predicted Wannier-centre configurations obtained using models trained on progressively more chemically relevant datasets. (a) Model trained without transition-metal dichalcogenides (TMDs), which fails to recover the non-atomic Wannier-centre distribution and localizes the predicted centres close to atomic positions. (b) Model trained on TMDs but excluding Mo- and S-containing materials, which captures partial features of the bonding environment but incorrectly predicts the Wannier-centre arrangement. (c) Model trained with Mo- and S-containing materials while excluding monolayer MoS$_2$ from the training set, accurately reproducing the reference Wannier-centre configuration. (d) Root mean squared error (RMSE) of Wannier-centre predictions for each training scenario.}
    \label{fig:mos2}
\end{figure*}

\section*{Discussion}
The ability to transfer across length scales, chemical environments, and previously unseen compounds is a central requirement for ML models of electronic structure. \textsc{WannDiffuse} meets these requirements for the quantum-mechanical position operator describing electronic localization. Trained on small reference cells, \textsc{WannDiffuse} accurately predicts WCCs in systems orders of magnitude larger, enabling spectroscopic simulations of liquids and amorphous materials at scales that are inaccessible to routine first-principles approaches. This scalability stems from the locality of the learned electronic representation and directly exploits the real-space Wannier-function framework. The universal model further demonstrates that this capability extends across 67 chemical species and is not limited to the first-order moment of the position operator (the WCCs), but also applies to its second-order moment.  More generally, the present framework can in principle predict arbitrary higher-order moment tensors of the position operator, provided that the corresponding reference data are available for training.
The high numerical accuracy on relatively small datasets also enables reliable and efficient  evaluation of response properties through finite-difference derivatives, including Born effective charges and piezoelectric tensors.

Today, dynamical properties are increasingly accessed through MD driven by MLIPs, whose training typically relies on self-consistent-field (SCF) first-principles calculations of representative supercell configurations. We therefore envision that, in most applications, \textsc{WannDiffuse} will be trained alongside a MLIP within the same active-learning workflow. In practice, this would require only an additional non-self-consistent-field (NSCF) calculation followed by Wannierization, or potentially Wannierization directly on top of the SCF calculation already performed for the MLIP training data. The additional cost of training \textsc{WannDiffuse} would thus be largely reduced to the Wannierization step itself, whose computational overhead is typically negligible compared with the underlying SCF first-principles calculations. Looking ahead, we plan to extend the present universal model toward a foundation model for electronic localization that, in analogy with emerging foundation models for MLIPs, can be efficiently fine-tuned to arbitrary target systems and levels of accuracy.
\newpage
\section*{Methods}
\subsection*{Graph Neural Network Architecture\label{sec:gnn}}
 A graph data structure consists of a finite set of vertices, called nodes, along with a set of pairs of these nodes, called edges. In an undirected graph, these pairs are unordered, whereas in a directed graph, the pairs are ordered. To construct a graph neural network (GNN) for atomistic systems, we first need to define the graph connectivity, that is, which atoms and WCCs (nodes) are connected to which neighbors (edges).
 We compute this connectivity using the matscipy library \cite{Grigorev2024}, which provides tools for atomistic simulations, including efficient routines for neighbor-list construction under periodic boundary conditions. In this case, we use matscipy to identify all pairs of atoms within a specified cutoff radius, returning indices of connected atom pairs and integer lattice shifts for periodic images.

We then construct the graph representation using jraph \cite{jraph2020github}, a library for implementing graph neural networks (GNNs) in \textsc{JAX}, developed by DeepMind. jraph provides a flexible GraphsTuple data structure for representing graphs, which stores node features, edge features, global attributes, and connectivity.
In our case, we create a graph object from the WCCs' positions, atomic positions, atomic species, and the unit cell. The nodes correspond to atoms and WCCs, while the edges are determined by the relative vectors within the user-defined cutoff radius. This representation forms the input to the GNN.

For our ML model, we employ NequIP. The architecture comprises two hidden layers, configured as:
\begin{gather} \label{eq:model}
    32\times0e + 32\times1o \\ \nonumber
    64\times0e + 64\times1o
\end{gather}
To ensure clarity regarding NequIP's notation, the basic structure of a layer is
represented by $\langle N\rangle\times\langle L\rangle\langle P\rangle$, where
$\langle N\rangle$ denotes the number of channels or independent copies of the
irreducible representation, $\langle L\rangle$ signifies the angular momentum
quantum number, and $\langle P\rangle$ indicates the parity (even `e' or odd~`o').

The output layer is configured as $1\times1o + 1\times0e$ per node,
simultaneously producing a polar displacement vector and a scalar spread for
each WCC. The two outputs are obtained by projecting the final layer
features onto their respective irreducible representations: the $1\times1o$
component gives the predicted WCC position, consistent with the odd
parity of a displacement under inversion, while the $1\times0e$ component gives
the predicted spread, which transforms as a scalar under all symmetry operations
and is therefore invariant to rotations and reflections. This joint output allows
the model to learn both the spatial location and the spatial extent of each
Wannier function within a single forward pass.

The network is trained using a loss function $\mathcal{L}$ based on the mean
squared difference between the true noise $\epsilon$ and the predicted noise
$\epsilon_\theta$,
\begin{gather} \label{eq:loss_model}
    \mathcal{L}(\theta)
    = \mathbb{E}_{x_0,\,\epsilon}
      \bigl[\|\epsilon_\theta(x_0 + \epsilon,\,\sigma) - \epsilon\|^2\bigr]  \\ \nonumber 
     + \lambda\, \mathbb{E}_{x_0}\bigl[\|\hat{s} - s_0\|^2\bigr],
\end{gather}

where the first term is the score-matching objective derived in
Sec. \SMdiff .3, the second term penalizes errors in the predicted
spreads $\hat{s}$ relative to the true spreads $s_0$, and $\lambda$ is a
weighting hyperparameter balancing the two objectives. 

\subsection*{GNN Model Training}

\begin{table*}[htbp] 
\centering
\begin{ruledtabular}
\begin{tabular}{lcccc}
 & \textbf{Liquid water} & \textbf{Amorphous Si} & \textbf{BaTiO$_3$} & \textbf{Universal} \\
No. of training structures            & 800 & 120 & 50   & 2807 \\
No. of validation structures & 200 & 30 & 10  & 100 \\
Diffusion timesteps          & 15   & 15  & 50   & 15 \\
Effective training samples   & 12{,}000 & 1{,}800 & 2{,}500 & 42{,}105 \\
Cutoff radius (\AA)          & 6.0  & 5.5 & 4.5  & 4.0 \\
Avg.\ neighbor count         & 220 & 110 & 65 & 360 \\
No.\ atomic species          & 2    & 1   & 3    & 67 \\
Species embedding dim.       & 16   & 16  & 4    & 128 \\
Initial learning rate        & $10^{-2}$ & $10^{-2}$ & $10^{-2}$ & $10^{-3}$ \\
Weight decay                 & $10^{-3}$ & $10^{-3}$ & $10^{-5}$ & $10^{-4}$ \\
LR schedule transition       & $1.5\times10^{4}$ & $1.5\times10^{4}$ & $1.5\times10^{4}$ & $3.5\times10^{4}$ \\
LR exponential decay factor  & 0.985 & 0.99 & 0.99 & 0.99 \\
\end{tabular}
\end{ruledtabular}
\caption{Dataset sizes, architectures, and optimization settings for \textsc{wanndiffuse} across all trained systems. All models use the AdamW optimizer with a stepwise exponential learning rate schedule. The number of atomic species reported here does not include WCCs, which are treated as an additional species in the graph and are always present, so the effective number of species is increased by one. }
\label{tab:training_details}
\end{table*}

\paragraph{Liquid water:}
For liquid water, we train \textsc{WannDiffuse} using the 32-molecule PIMD dataset of Grisafi et al.\ \cite{Grisafi18}. Dataset, architecture, and optimization settings are summarized in Table~\ref{tab:training_details}.

\paragraph{Amorphous silicon:}
For amorphous silicon, we train \textsc{WannDiffuse} on structures generated using the data augmentation protocol of the AiiDA-TrainsPot automated workflow~\cite{Bidoggia2026}.
Starting from a set of 13 crystalline Si structures, additional configurations were generated by cell replication with a minimum cell length of 24~\AA\ and up to 120 atoms per cell, combined with random atomic-position rattling up to 30\% of the nearest-neighbor distance, lattice strain ranging from $-20\%$ to $+30\%$, and vacancy defects introduced in 20\% of the augmented structures at 2 vacancies per structure. Dataset sizes, architectures, and optimization settings are summarized in Table~\ref{tab:training_details}.

\paragraph{BaTiO$_3$:}
Also for BaTiO$_3$, we train \textsc{WannDiffuse} using the dataset augmentation protocol of AiiDA-TrainsPot~\cite{Bidoggia2026}. Starting from 2 input structures (cubic and tetragonal phases), additional configurations were generated by cell replication with a minimum cell length of 6~\AA\ and up to 50 atoms per cell, combined with random atomic-position rattling up to 30\% of the nearest-neighbor distance and lattice strain ranging from $-30\%$ to $+30\%$. 
This yielded 25 augmented configurations per input structure, for a total of 50 training structures. Dataset sizes, architectures, and optimization settings are summarized in Table~\ref{tab:training_details}.

\paragraph{Universal model:}
The universal model is trained on a large dataset of Wannier functions describing the occupied bands of insulating materials spanning diverse chemistries and structures obtained in Ref. \cite{JFdataset26} 
drawn from a 67-species chemical space to ensure broad elemental coverage. Dataset sizes, architectures, and optimization settings are summarized in Table~\ref{tab:training_details}. The joint output formulation enables the model to simultaneously predict the denoising direction for WCC positions at each diffusion timestep and the associated spreads. The predicted spreads are invariant to rotations and reflections by construction, consistent with the equivariant structure of the network.

\subsection*{IR spectroscopy}
The total dipole moment $\mu$ of the system is computed as $\mu = \sum_i \mu_i$, where $\mu_i$ is the molecular dipole moment of the $i^{\rm {th}}$ water molecule. Each $\mu_i$ is determined from the positions of the nuclei and the corresponding WCCs, which are predicted using the E(3)-equivariant graph neural network diffusion model described in Sec.~\ref{sec:gnn}. 
Ref. \cite{bader94} provides a derivation using Fermi's golden rule and the Poynting vector with a classical approach. The IR spectra can be computed as:

\begin{eqnarray}\label{eq:ir}
   I(\omega) &=& \frac{\pi \beta}{3cV} \omega^2 \int_{-\infty}^{\infty}  \braket{\vec{\mu} (0) | \vec{\mu} (t) } \rm{e}^{ -i \omega t} dt \\ \nonumber
    &=& \frac{\pi \beta \omega^2}{3cV} \mathcal{F}[ \braket{ \vec{\mu} (0)  | \vec{\mu} (t) }]
\end{eqnarray}

Here,  $\beta$ = $(k_B T)^{-1}$, where $k_B$ is the Boltzmann constant and $T$ is the temperature. $\mathcal{F}$ is the Fourier transform, $V$ denotes the volume of the simulation cell, $c$ is the speed of light, and $\omega$ is the frequency.

\subsection*{First-principles simulations}
DFT calculations for liquid water, amorphous Silicon and Barium titanate were performed using \textsc{Quantum ESPRESSO} \cite{Giannozzi2009, Giannozzi2017, giannozzi2020quantum} in conjunction with the SSSP Efficiency pseudopotential library (version 1.3, PBEsol functional) \cite{Prandini2018, Blochl1994, DalCorso2014, Hamann2013, Vanderbilt1990, Garrity2014, Kucukbenli2014}. High-throughput calculations were orchestrated with the AiiDA workflow infrastructure \cite{Pizzi16, Huber20}, using the \texttt{aiida-wannier90-workflows} package (\url{https://github.com/aiidateam/aiida-wannier90-workflows}), which automates the submission of \textsc{Quantum ESPRESSO} and \textsc{Wannier90} \cite{Pizzi16, Pizzi2020} calculations to remote computing clusters, handles the parsing and storage of results in a centralized database, and orchestrates the sequential execution of all simulations. 
Brillouin-zone sampling was generated using the “precise” protocol of the AiiDA PW WorkChain \cite{deMirandaNascimento2026}.
Each workflow takes a crystal structure as input and produces, as outputs, the Wannier-interpolated band structure, the real-space MLWFs, and the corresponding WCCs. For these calculations with \texttt{aiida-wannier90-workflows}, the electronic type was set to insulator, and the selected columns of the density matrix (SCDM) method \cite{Damle17} was employed for the initial projections. 

For an insulator, the number of occupied bands and the number of Wannier functions to compute are fixed by the total valence-electron count of the constituent atoms, since each doubly-occupied MLWF accounts for two electrons. The number of computed bands was set equal to this number. Applying this rule with the valence configurations defined in the SSSP PBEsol library gives for primitive cells 128 Wannier functions for the 32-molecule liquid-water system, 128 for the 64-atom amorphous Si system,  9 for MoS$_2$ and 20 for BaTiO$_3$.

Amorphous silicon structures from first-principles were generated by melt-quench MD starting from a 64-atom crystalline Si cell, using \textsc{Quantum ESPRESSO} with the PBEsol exchange-correlation functional, a plane-wave cutoff of 35~Ry (280~Ry for the charge density), Methfessel-Paxton smearing with a broadening of 0.02~Ry. The cell was first melted at 3500~K using Verlet dynamics with velocity rescaling (tolerance 100~K) and a timestep of 1~fs. The system was subsequently quenched in stages at 1500~K, 800~K, 500~K, and 300~K, with 100 MD steps performed at each intermediate temperature. At the final temperature of 300~K, the trajectory was extended to generate a total of 5,000 structures for use in dataset augmentation. 
From these, 300 uncorrelated structures were selected to compute the radial distribution function.

For the universal model, we used projectability-disentangled Wannier functions (PDWFs) \cite{Qiao2023a} to obtain MLWFs \cite{marzari_prb_1997} for the valence plus low-lying conduction bands, which were then partitioned into separate valence and conduction sets using manifold-remixed Wannier functions (MRWFs) \cite{Qiao2023b}. 
The PDWFs were obtained using the \textsc{wannier90} code \cite{Pizzi2020}, while the MRWFs were obtained using the \texttt{Wannier.jl} code (\url{https://github.com/qiaojunfeng/Wannier.jl}). Both codes were orchestrated through the AiiDA \cite{Huber20} and \texttt{aiida-wannier90-workflows} infrastructure described above. 
With this pipeline, we obtained valence-only Wannier functions for 5032 insulators selected from the Materials Cloud three-dimensional structure database (MC3D) \cite{huber_mc3d_2025}. Of these, we retained 2907 structures for which all semicore states were fully Wannierized, since the diffusion model requires a constant number of WCCs per chemical species. 
The centers of these valence-only Wannier functions form the starting dataset for the diffusion model.

\subsection*{Polarization-derived response properties}

To evaluate the ability of \textsc{WannDiffuse} to reproduce polarization-related response properties, we applied a model trained on rattled BaTiO$_3$ configurations and computed polarization derivatives from the predicted WCCs using finite-difference approaches.

\paragraph*{Born effective charges}
For each atom $\kappa$, a small
displacement $\Delta u_{\kappa,\beta}$ was applied to that atom
along Cartesian direction $\beta$ (of magnitude 0.01~\AA), with all other atomic positions and the lattice vectors held fixed. At each displaced configuration the WCCs were predicted by the model, and the total polarization,
$\mathbf{P}$ -- ionic plus electronic, Eq.~(\ref{eq:wannier_polarization}), using the actual
displaced ionic positions $R_I$ -- was evaluated from the predicted
WCCs together with the nominal ionic charges $Z_I$. Because
Eq.~(\ref{eq:wannier_polarization}) is defined modulo the polarization quantum
$e\mathbf{R}/V_c$, for each pair of displaced structures the branch
of $P_\alpha(\pm\Delta u_{\kappa,\beta})$ closest to the
polarization of the undisplaced reference structure was selected
before differencing, so that finite differences are not affected by
polarization-quantum jumps. Born effective charges were then
obtained from finite differences of the total polarization,
\begin{equation}
Z^*_{\kappa,\alpha\beta} = \frac{\Omega}{e}
\frac{\partial P_\alpha}{\partial u_{\kappa,\beta}} \approx
\frac{\Omega}{e}
\frac{P_\alpha(\Delta u_{\kappa,\beta}) - P_\alpha(-\Delta u_{\kappa,\beta})}
{2\,\Delta u_{\kappa,\beta}},
\label{eq:bec}
\end{equation}
where $\Omega$ is the unit-cell volume, $e$ is the elementary
charge, and $P_\alpha$ is the $\alpha$-component of the total
(ionic + electronic) polarization defined in Eq.~(\ref{eq:wannier_polarization}).

\paragraph*{Spontaneous polarization}
Experimental atomic coordinates for
the tetragonal ferroelectric phase were obtained from the ICSD
~\cite{Megaw1962}. 
The electronic polarization was computed directly from the predicted WCCs following the modern theory of polarization \cite{Resta1994}; the
total polarization used below is the sum of this electronic term
and the ionic term $\sum_I e Z_I R_I / V_c$ evaluated at the
corresponding ionic positions (Eq.~(\ref{eq:wannier_polarization})). The spontaneous polarization was evaluated as
\begin{equation}
\Delta P = P_{\text{ferro}} - P_{\text{centro}},
\label{eq:deltap}
\end{equation}
where $P_{\text{centro}}$ is the total polarization of the
centrosymmetric cubic reference structure, with the branch of
$P_{\text{ferro}}$ chosen as the one continuously connected to
$P_{\text{centro}}$ along the cubic-to-tetragonal distortion path.

\paragraph*{Clamped-ion piezoelectric tensor}
 The electronic (clamped-ion)
piezoelectric tensor was computed by finite differences of the
polarization with respect to homogeneous strain. Strained lattice
vectors were generated by applying positive and negative Voigt
strain perturbations $\pm\delta\eta_j$ (of magnitude 1\%) to the equilibrium cell, with ionic positions held
fixed in fractional coordinates. For normal strains ($xx$, $yy$,
$zz$), the lattice vectors were scaled along the corresponding
Cartesian direction, while for shear strains ($yz$, $xz$, $xy$)
symmetric engineering strain tensors were employed. At each strained
configuration the total polarization $P_\alpha$ (ionic + electronic,
Eq.~(\ref{eq:wannier_polarization}), with ionic positions $R_I = H_\pm \cdot$(fractional coordinates)) was evaluated from the predicted WCCs, again selecting the branch continuous with the unstrained reference before
differencing. 
The piezoelectric coefficients were then evaluated as
\begin{equation}
e_{\alpha\beta} = \left.\frac{\partial P_\alpha}{\partial \eta_\beta}
\right|_{\text{ions fixed}} \approx
\frac{P_\alpha(+\delta\eta_\beta) - P_\alpha(-\delta\eta_\beta)}
{2\,\delta\eta_\beta},
\label{eq:piezo}
\end{equation}
where $\eta_\beta$ is the Voigt-notation strain component and
$P_\alpha$ is again evaluated from the total, \textsc{WannDiffuse}-predicted
polarization. For each strain component, the equilibrium cell matrix
$H_0$ is deformed according to $H_\pm = H_0 (I + \delta \eta^{\pm})$,
where $\eta^{\pm}$ is the symmetric strain matrix corresponding to a
Voigt component evaluated at $+\delta$ and $-\delta$. For shear
strains (e.g.\ $yz$, $xz$, $xy$), the off-diagonal entries of
$\eta$ carry a factor of $1/2$ in accordance with the Voigt
convention. Ionic positions are held fixed in fractional coordinates
under the deformation, isolating the clamped-ion (electronic)
response.

\section*{Acknowledgements}
R.D.M. acknowledges support from a PhD scholarship co-funded by the Italian Ministry of University and Research (MUR) under D.M. 630/2024 and by ENI S.p.A, within the National Recovery and Resilience Plan (PNRR), Mission 4, Component 2, Investment 3.3, funded by the European Union -- NextGenerationEU.
A. M. and Na. M. acknowledge partial support from the European Commission through the Centre of Excellence “MaX - Materials Design at the Exascale” (HORIZON-EUROHPC, Grant No. 101093374). 
R.D'S would like to thank Florian Pabst for providing the MD data for the 512-molecule liquid-water system from Ref. \cite{Pabst2025}, 
as well as for useful discussions.
The views and opinions expressed are solely those of the authors and do not necessarily reflect those of the European Union, nor can the European Union be held responsible for them.

\section*{Data Availability}
All data and code associated with this work will be made publicly available. The source code for \textsc{WannDiffuse} will be released under the MIT license on GitHub at \texttt{github.com/ARGO-SISSA/WannDiffuse}. The repository is thoroughly documented and includes worked examples covering single-material calculations, universal models with and without Wannier-function spread predictions, and three-dimensional visualization tools for Wannier-function diffusion. The datasets supporting the findings of this study will be made openly available through Materials Cloud~\cite{MC_dataset} upon publication.
\section*{Author contributions}
A.M. conceived and coordinated the project and supervised its overall scientific development, with contributions from A.Z. and Ni.M. A.Z. proposed the diffusion-based approach, while Ni.M. identified amorphous silicon as a key test case. R.D'S. developed \textsc{WannDiffuse} and led the training and validation campaigns under the supervision of A.M. and A.Z. Na.M. developed the AiiDA workflow to accelerate training and carried out part of the simulations used to generate the training data. R.D.M. calculated the piezoelectric tensors and contributed to testing \textsc{WannDiffuse}. J.Q. provided the training data for the universal model. R.D'S. prepared all figures and, together with A.M., wrote the first draft of the manuscript. All authors contributed to the analysis and interpretation of the results and to the revision of the final manuscript.

%

\end{document}